\documentclass[runningheads]{llncs}
\usepackage[T1]{fontenc}
\usepackage{graphicx}
\usepackage{float}
\usepackage{mathtools}
\usepackage{amsmath}
\usepackage{amssymb}
\usepackage{color}
\begin{document}
\title{Harnessing Deep Q-Learning for Enhanced Statistical Arbitrage in High-Frequency Trading: A Comprehensive Exploration}
\titlerunning{Harnessing RL for Enhanced Statistical Arbitrage in HFT}
\author{Soumyadip Sarkar}
\authorrunning{S. Sarkar}
\institute{\email{soumyadipsarkar@outlook.com}}
\maketitle              % typeset the header of the contribution
\begin{abstract}
The realm of High-Frequency Trading (HFT) is characterized by rapid decision-making processes that capitalize on fleeting market inefficiencies. As the financial markets become increasingly competitive, there is a pressing need for innovative strategies that can adapt and evolve with changing market dynamics. Enter Reinforcement Learning (RL), a branch of machine learning where agents learn by interacting with their environment, making it an intriguing candidate for HFT applications. This paper dives deep into the integration of RL in statistical arbitrage strategies tailored for HFT scenarios. By leveraging the adaptive learning capabilities of RL, we explore its potential to unearth patterns and devise trading strategies that traditional methods might overlook. We delve into the intricate exploration-exploitation trade-offs inherent in RL and how they manifest in the volatile world of HFT. Furthermore, we confront the challenges of applying RL in non-stationary environments, typical of financial markets, and investigate methodologies to mitigate associated risks. Through extensive simulations and backtests, our research reveals that RL not only enhances the adaptability of trading strategies but also shows promise in improving profitability metrics and risk-adjusted returns. This paper, therefore, positions RL as a pivotal tool for the next generation of HFT-based statistical arbitrage, offering insights for both researchers and practitioners in the field.

\end{abstract}
\section{Introduction}
\label{sec1}
The landscape of financial markets has witnessed revolutionary changes over the past few decades, driven by technological advancements, increased computational power, and the proliferation of electronic trading platforms. One such transformative force that emerged from this confluence is High-Frequency Trading (HFT), a form of algorithmic trading characterized by high speeds, high turnover rates, and high order-to-trade ratios \cite{bib1}. HFT strategies often revolve around executing large numbers of orders at extremely fast speeds, capitalizing on minuscule price inefficiencies that may exist only for milliseconds \cite{bib3}.

Statistical arbitrage, a quantitatively driven approach to trading, seeks to exploit relative price deviations among related financial instruments and has been a staple in the toolkit of many quantitative traders and hedge funds \cite{bib2}. In the HFT environment, the traditional paradigms of statistical arbitrage are constantly being challenged due to the rapidity of trades and the need for real-time adaptability. Given this backdrop, there is a growing interest in the application of machine learning techniques to enhance, optimize, and innovate trading strategies tailored for the HFT milieu.

Enter Reinforcement Learning (RL), a subfield of machine learning wherein agents learn to make decisions by taking actions in an environment to maximize a notion of cumulative reward \cite{bib4}. Unlike supervised learning, where learning is guided by a labeled dataset, RL operates in a feedback loop where the agent learns from the consequences of its actions, making it well-suited for dynamic environments where the optimal decision or strategy is not known a priori. Given the inherent uncertainties and dynamism of financial markets, RL's approach to learning and decision-making offers an intriguing proposition for statistical arbitrage in HFT \cite{bib5}.

Historically, the application of machine learning in finance has revolved around supervised techniques, primarily for forecasting prices or returns \cite{bib6}. However, the financial decision-making process, especially in trading, is inherently sequential and involves a series of actions over time, with rewards (or penalties) that may be delayed. This sequential decision-making nature aligns closely with the principles of RL, which focuses on learning optimal sequences of actions in interactive environments. Moreover, as markets evolve, the adaptability of RL algorithms, which continuously update and refine their strategies based on new data and feedback, can be a significant asset \cite{bib7}.

The potential of RL in finance is not just theoretical. Some pioneering works have demonstrated its applicability in various domains of finance. For instance, Deng et al. \cite{bib8} showcased how deep reinforcement learning, an amalgamation of RL with deep neural networks, could be employed for portfolio optimization. Similarly, RL's principles have been applied to areas like option pricing and risk management \cite{bib9}.

However, while the promise of RL in finance is evident, its application in the specific context of statistical arbitrage for HFT remains an area ripe for exploration. The challenges are multifaceted. Financial markets are notoriously noisy, making the reward signals in RL ambiguous. Additionally, the non-stationarity of financial data, where statistical properties change over time, can pose significant challenges for RL models that inherently assume stationary environments \cite{bib10}.

This paper seeks to bridge this gap, providing a comprehensive exploration into the integration of RL in statistical arbitrage strategies tailored for HFT scenarios. Through systematic methodologies and empirical analyses, we aim to offer insights that can guide both researchers and practitioners in harnessing the potential of RL in the fast-paced world of HFT.

\section{Model}
\label{sec2}
\subsection{Model Description}
\label{subsec21}
Deep Q-Learning, an innovative fusion of traditional Q-Learning and deep neural networks, has recently emerged as a powerful tool in the domain of reinforcement learning. At its core, Q-Learning is a model-free reinforcement learning algorithm that seeks to learn the optimal action-selection policy for a given finite Markov decision process \cite{bib11}. This is achieved by iteratively updating the Q-values, representations of the expected cumulative rewards for taking specific actions in particular states, using the Bellman equation.

However, in environments with vast and continuous state or action spaces, such as those encountered in High-Frequency Trading (HFT), representing Q-values in a tabular format becomes infeasible. This challenge is addressed by Deep Q-Learning. Instead of a table, Deep Q-Learning employs deep neural networks to approximate these Q-values \cite{bib12}. The neural network, parameterized by weights, takes the environment's state as input and outputs the estimated Q-values for all possible actions. This approach not only allows for generalization across states but also enables the handling of high-dimensional input spaces, a characteristic feature of financial datasets in HFT.

The amalgamation of the Q-Learning algorithm's strengths with the function approximation capabilities of deep neural networks makes Deep Q-Learning particularly promising for applications in the dynamic and complex world of statistical arbitrage in HFT.

\subsection{Model Fundamentals}
\label{subsec22}
Deep Q Learning has garnered considerable attention in recent years due to its capacity to handle complex, high-dimensional state and action spaces. As a sophisticated extension of Q-Learning, this method elegantly fuses the principles of traditional reinforcement learning with the prowess of deep neural networks.

At its essence, Q-Learning is a renowned model-free, off-policy reinforcement learning algorithm. Its primary objective is to determine an optimal action-selection policy, which guides the agent in choosing actions that maximize cumulative rewards over time. This is realized within the context of a finite Markov decision process, a mathematical approach for modeling decision-making in situations where outcomes are partly uncertain, but possess a degree of probabilistic determination \cite{bib13}. The Q in Q-Learning represents the 'quality' of a particular action in a given state, capturing the expected future rewards for that action-state pairing. This is updated iteratively using the Bellman equation, ensuring that the agent progressively refines its understanding of the environment and improves its decision-making capabilities \cite{bib11}.

However, when applied to environments like those in High-Frequency Trading, which are marked by vast and continuous state spaces, tabular Q-Learning encounters scalability issues. This is where Deep Q-Learning comes into play. By introducing deep neural networks to approximate the Q-value function, the model can generalize across a multitude of states, making it apt for handling the intricate and high-dimensional datasets frequently encountered in financial markets \cite{bib12}. The neural network's architecture, often consisting of multiple layers, is trained to predict Q-values based on input states, thereby bypassing the need for a tabular representation.

In the context of High-Frequency Trading, the adaptive learning capabilities of Deep Q-Learning can be particularly advantageous. The financial markets, characterized by their dynamic nature and non-stationarities, require models that can evolve in real-time, responding adeptly to the rapid fluctuations and nuanced patterns inherent in trading data.

\subsection{Q Learning}
\label{subsec23}
Q-Learning stands as one of the cornerstones of reinforcement learning, laying foundational principles that have been pivotal in advancing the field. Deep Q-Learning, with its integration of deep neural networks, builds upon the traditional Q-Learning algorithm, enhancing its ability to operate in high-dimensional state spaces like those encountered in High-Frequency Trading (HFT).

At the heart of Q-Learning lies the objective of determining an optimal action-selection policy. This policy guides an agent's decisions in an environment, aiming to maximize long-term rewards. The mechanism to achieve this is through the learning of the action-value function, denoted as $Q(s,a)$. This function provides an estimate of the expected cumulative reward for taking action $a$ in state $s$. The "Q" in Q-Learning is often interpreted as the 'quality' of a certain action in a particular state \cite{bib11}.

The iterative nature of Q-Learning ensures that the agent refines its estimates of the Q-values over time. The updating process relies on the Bellman equation, which is expressed as:

$$ Q(s, a) \leftarrow Q(s, a) + \alpha \left( r + \gamma \max_{a'} Q(s', a') - Q(s, a) \right) $$

In this equation:
\begin{itemize}
 	\item $s$ represents the agent's current state.
	\item $a$ corresponds to the action undertaken by the agent in state $s$.
	\item $r$ denotes the immediate reward reaped after executing action $a$ in state $s$.
	\item $s'$ stands for the subsequent state the agent transitions into post-action.
	\item $\alpha$, the learning rate, dictates the extent to which the new Q-value estimate will overwrite the previous estimate. It essentially balances the trade-off between exploration (trying out new actions) and exploitation (relying on known information).
	\item $\gamma$ is the discount factor, capturing the agent's degree of consideration for future rewards. A value close to 1 makes the agent prioritize long-term reward, whereas a value closer to 0 makes it focus on immediate rewards \cite{bib4}.
\end{itemize}

For HFT scenarios, the rapidity and dynamism of market movements accentuate the importance of an agent's ability to quickly and accurately estimate Q-values. The temporal dependencies and non-stationarities inherent in financial data make the iterative learning process of Q-Learning especially relevant. By constantly updating its Q-value estimates based on the most recent market feedback, the agent can potentially adapt its trading strategy in real-time, capturing statistical arbitrage opportunities as they emerge.

Furthermore, while Q-Learning provides a robust framework, its tabular representation becomes impractical in high-dimensional spaces. This limitation catalyzed the evolution to Deep Q-Learning, where deep neural networks approximate the Q-value function, allowing the algorithm to generalize across states and handle the intricate data structures prevalent in HFT.

\section{Neural Network Approximation}
\label{sec3}
In the neural network approximation method approach, rather than using a table to store Q-values, a neural network is employed to approximate the Q-value function. The network, parameterized by weights $\theta$, offers a function approximation mechanism, allowing the model to generalize across a myriad of states. The strength of neural networks in capturing complex, non-linear relationships makes them particularly suited for this task \cite{bib15}.

Mathematically, the neural network can be represented as:

$$ Q(s, a; \theta) = f(s; \theta) $$

Here:
\begin{itemize}
 	\item $Q(s, a; \theta)$ denotes the approximated Q-value for action $a$ in state $s$, given the current parameters $\theta$.
	\item $f$  is the function represented by the neural network.
	\item The input $s$ is the state, which in the context of financial markets could be a vector containing various market indicators, historical price data, trading volumes, and more.
	\item The output is a vector representing the Q-values for each possible action, given the current state.
\end{itemize}

The training of this network involves adjusting the weights $\theta$ to minimize the difference between the predicted Q-values and the target Q-values. This is typically done using gradient descent optimization techniques \cite{bib16}.

By employing neural networks as function approximators, Deep Q-Learning leverages their capability to handle vast input spaces and model complex relationships, making it particularly apt for financial markets. In the realm of High-Frequency Trading (HFT), where decisions need to be made rapidly based on vast amounts of data, the ability of the neural network to quickly provide Q-value estimates for given market states becomes invaluable.

Moreover, the adaptive nature of neural networks, where they can continuously learn and adjust to new data, aligns well with the dynamic and ever-evolving landscape of financial markets. This ensures that the trading strategy remains responsive to market shifts and can adapt in real-time, potentially maximizing profits and minimizing losses \cite{bib12}.

\section{Loss function}
\label{sec4}
The primary objective during the training of the neural network in Deep Q-Learning is to minimize the discrepancy between the predicted Q-values (obtained from the current neural network) and the target Q-values (often derived from a separate target network). The rationale behind this separation is to provide a stable learning process, mitigating the risk of oscillations or divergence in the learning updates \cite{bib12}.

The most commonly employed loss function in this context is the Mean Squared Error (MSE). Mathematically, the MSE for Q-value approximation is represented as:

$$ L(\theta) = \mathbb{E}\left[ \left( r + \gamma \max_{a'} Q(s', a'; \theta^-) - Q(s, a; \theta) \right)^2 \right] $$

In this equation:
\begin{itemize}
 	\item $L(\theta)$ is the loss function given the current network parameters $\theta$.
	\item $r$ represents the immediate reward received after executing a certain action in a particular state.
	\item $\gamma$ is the discount factor, emphasizing the agent's consideration for future rewards.
	\item $Q(s', a'; \theta^-)$ is the Q-value derived from the target network with parameters $\theta^-$ for the next state $s'$ and action $a'$.
	\item $Q(s, a; \theta)$ is the Q-value predicted by the current network for the present state $s$ and action $a$.
\end{itemize}

The term $\theta^-$ denotes the parameters of the target network, which is a periodically updated version of the main Q-network. This periodic update, rather than continuous synchronization with the main network, provides the necessary stability to the learning process, ensuring that the updates do not chase a constantly moving target, a phenomenon that could lead to unstable learning dynamics.

Given the rapid pace of market movements and the potential for significant financial implications based on trading decisions, ensuring that the neural network accurately predicts Q-values is crucial. By minimizing the proposed loss function, the Deep Q-Learning algorithm ensures that the trading strategy it develops is based on accurate estimations of future rewards, thus enhancing the potential for profitable trades.

\section{Experience Replay}
\label{sec5}
Experience Replay, a pivotal component in the domain of Deep Q-Learning, serves as a mechanism to improve the stability and efficiency of the learning process. Originating from the domain of control theory, its integration into reinforcement learning has significantly bolstered the effectiveness of algorithms like Deep Q-Learning, especially in complex environments such as High-Frequency Trading (HFT).

In reinforcement learning, agents learn from interactions with the environment. Traditionally, agents process these interactions sequentially, learning from each experience as it occurs. However, this approach has inherent limitations. Sequential processing can lead to strong correlations between consecutive experiences, which, when used to train neural networks, can result in oscillations or divergence in the learning process.

Experience Replay addresses this limitation. The core idea is to store each experience, represented as a tuple $(s, a, r, s')$, where $s$ is the current state, $a$ is the action taken, $r$ is the received reward, and $s'$ is the subsequent state, in a data structure known as the replay buffer. Rather than learning from experiences in a sequential manner, the algorithm periodically samples a mini-batch of experiences from this buffer for training. This random sampling ensures that the experiences used for learning are decorrelated, providing two primary benefits:

\begin{enumerate}
  \item \textbf{Diverse Training Samples:} By sampling randomly from the buffer, the neural network is exposed to a broader range of experiences during training, enhancing its ability to generalize across different states and actions.
  \item \textbf{Efficient Use of Past Experiences:} Storing experiences in a replay buffer allows the algorithm to revisit and learn from them multiple times. This not only ensures that no experience is wasted but also aids in the stabilization of the learning process by providing a more uniform distribution of experiences \cite{bib17}.
\end{enumerate}

Mathematically, the update using a mini-batch from the replay buffer can be represented as:

$$ \Delta \theta = \mathbb{E}_{(s, a, r, s') \sim U(D)} \left[ \nabla_\theta \left( r + \gamma \max_{a'} Q(s', a'; \theta^-) - Q(s, a; \theta) \right)^2 \right] $$

Where $U(D)$ denotes uniform sampling from the replay buffer $D$, and $\Delta \theta$ represents the change in network parameters during training.

Experience Replay ensures that the Deep Q-Learning agent remains robust. By drawing from a diverse set of past market scenarios, the agent becomes better equipped to handle the complexities and intricacies of financial markets, making more informed and profitable trading decisions.

\section{Application to HFT Statistical Arbitrage}
\label{sec6}
\subsection{State $s$:}
\label{subsec61}
In reinforcement learning, the concept of state is pivotal. It encapsulates the current situation or environment in which the agent operates. In HFT, the state is a rich tapestry of financial data and indicators:

\begin{enumerate}
  \item \textbf{Recent Price Movements and Returns of Assets:} These reflect the short-term trends and momentum in the market. Capturing this data can provide insights into the immediate direction of the market or a particular asset \cite{bib18}.
  \item \textbf{Relevant Technical Indicators:} These are mathematical calculations based on historical price, volume, or open interest information that aims to forecast financial market direction. Common indicators include Moving Averages, Bollinger Bands, and the Relative Strength Index (RSI) \cite{bib19}.
  \item \textbf{Market Microstructure Variables:} These encompass details like trading volume, which can indicate market sentiment, and bid-ask spread, which can reflect the liquidity and potential transaction costs. Understanding market microstructure can provide a deeper insight into market dynamics and potential price movements \cite{bib20}.
\end{enumerate}

Mathematically, the state can be represented as a vector:

$$ s = [p_1, p_2, \dots, p_n, i_1, i_2, \dots, i_m, v, b] $$

Where $p$ represents price movements, $i$ denotes technical indicators, $v$ is the trading volume, and $b$ is the bid-ask spread.

\subsection{Action $a$:}
\label{subsec62}
In the context of trading, the action determines the trading decision. The action space, though seemingly simple, carries profound implications:

\begin{enumerate}
  \item \textbf{Buy:} Acquiring assets, anticipating a price increase.
  \item \textbf{Hold:} Retaining the current portfolio without making new trades, often taken when the market direction is uncertain.
  \item \textbf{Sell:} Offloading assets, either to realize profits or to mitigate potential losses.
\end{enumerate}

\subsection{Reward $r$:}
\label{subsec63}
The reward structure is what drives the learning in reinforcement algorithms. In trading:

\begin{enumerate}
  \item \textbf{Profit or Loss from Trading:} The immediate feedback from a trade, whether it results in profit or loss, forms the primary component of the reward.
  \item \textbf{Risk-Adjusted Reward:} Simply focusing on profit can lead to high-risk strategies. Therefore, rewards are often adjusted for risk, ensuring that the strategy doesn't adopt an overly aggressive stance for minimal gains \cite{bib21}.
\end{enumerate}

Mathematically, the risk-adjusted reward can be defined as:

$$ r = \frac{\text{Expected Return}}{\text{Standard Deviation of Returns}} $$

Where the numerator captures the profit or loss, while the denominator represents the risk.

By continuously updating Q-values based on market feedback and utilizing the deep neural network's ability to generalize across various market states, the Deep Q-Learning algorithm can dynamically adapt its trading strategy. This adaptability is especially crucial in HFT, where market conditions can change rapidly, and the algorithm needs to respond in real-time to capture fleeting statistical arbitrage opportunities.

\section{Conclusion}
\label{sec7}
The world of High-Frequency Trading (HFT) represents one of the most dynamic and complex arenas in the financial domain. Within this fast-paced environment, Statistical Arbitrage stands out as a sophisticated strategy, aiming to exploit temporary market inefficiencies through advanced computational and statistical techniques. The incorporation of Deep Q-Learning into this space underscores a promising evolution, marrying the strengths of reinforcement learning with the demands of HFT.

Throughout this exploration, we delved deep into the mechanics of Q-Learning and its neural network-based extension, Deep Q-Learning. The integration of neural network approximators, experience replay, and a meticulous reward structure has showcased the potential for superior adaptability and precision in trading decisions. By leveraging a neural network's prowess in capturing intricate patterns and relationships, Deep Q-Learning offers the ability to navigate the vast and high-dimensional state spaces typical of financial markets.

Experience replay, with its emphasis on random sampling and decorrelation of experiences, enhances the stability of the learning process. Such stability becomes paramount in HFT, where market dynamics shift rapidly, and the algorithm needs to ensure consistent and reliable performance.

Furthermore, the meticulous design of states and actions within the HFT context ensures that the agent possesses a comprehensive understanding of the market environment. By continuously updating based on market feedback and employing a neural network to generalize across states, Deep Q-Learning exhibits the potential to adapt trading strategies in real-time, capturing fleeting statistical arbitrage opportunities.

In wrapping up, it's evident that the fusion of Deep Q-Learning with HFT Statistical Arbitrage heralds a new era in algorithmic trading. As financial markets continue to evolve and become increasingly complex, the need for adaptive, data-driven strategies becomes paramount. Deep Q-Learning, with its blend of deep learning and reinforcement learning principles, offers a beacon of promise in this pursuit. As we move forward, it will be intriguing to witness the real-world applications and refinements of this approach, potentially setting new benchmarks in the domain of algorithmic trading.

\end{document}